\documentclass{svproc}
\usepackage{url}
\usepackage{graphicx}% Include figure files

\begin{document}
\def\vc#1{\mbox{\boldmath $#1$}}
\mainmatter              % start of a contribution
\title{Approximate sum rule for the electric dipole moment of light nuclei}
\titlerunning{Approximate sum rule for the light nuclear EDM}  % abbreviated title (for running head)
\author{Nodoka Yamanaka}
\authorrunning{Nodoka Yamanaka} % abbreviated author list (for running head)
\institute{IPNO, Universit\'{e} Paris-Sud, CNRS/IN2P3, F-91406, Orsay, France,\\
\email{yamanaka@ipno.in2p3.fr}
}

\maketitle              % typeset the title of the contribution

\begin{abstract}
The measurement of the electric dipole moment (EDM) is an excellent test of the standard model of particle physics, and the detection of a finite value is signal of a new source of CP violation beyond it.
Among systems for which the EDM can be measured, light nuclei are particularly interesting due to their high sensitivity to new physics.
In this proceedings contribution, we examine the sensitivity of the EDM of several light nuclei to the CP-odd one pion-exchange nucleon-nucleon interaction within the cluster model.
We suggest an approximate sum rule for the nuclear EDM.
\end{abstract}

\section{Motivation}

The electric dipole moment (EDM) \cite{engeledmreview,yamanakabook,atomicedmreview,chuppreview} is a good probe of CP violation.
One of the most notable point is the almost negligible standard model contribution \cite{pospelovckm,seng,smnuclearedm,smdeuteronedm,sm9beedm}.
Recently, the experimental measurement of the EDM using storage rings is being developed \cite{jedi}.
Here we discuss the EDM of light nuclei \cite{yamanakanuclearedmreview,devriesedmreview} as potentially interesting observables.

In the next section, we introduce the model and the interactions used in our work.
In Section \ref{sec:nuclearedm}, we show our results of the calculations of the EDM of light nuclei, from which an interesting counting rule is suggested.
The final section gives the summary.

\section{The model setup}

We consider the nucleons, the $\alpha$ ($^4$He), and triton ($^3$H) clusters as degrees of freedom, which are interacting themselves through phenomenological potentials \cite{av18,kanada,hasegawa,nishioka,schmid,yamada11b,yamada13c}.
The effect of the antisymmetrization is included using the orthogonality condition model \cite{saito}.

\begin{figure}
\begin{center}
\includegraphics[width=0.6\textwidth]{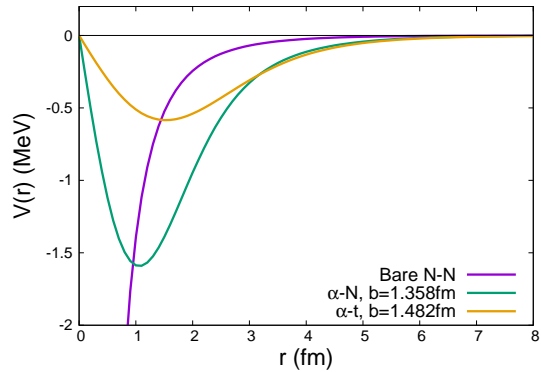}
\caption{
The shape of the CP-odd potentials.
}
\label{fig:folding}
\end{center}
\end{figure}

The CP-odd nuclear force is modeled by the one-pion exchange \cite{pvcpvhamiltonian3}:
\begin{eqnarray}
H_{P\hspace{-.45em}/\, T\hspace{-.5em}/\, }^\pi
& = &
\bigg\{ 
\bar{G}_{\pi}^{(0)}\,{\vc{\tau}}_{1}\cdot {\vc{\tau}}_{2}\, {\vc{\sigma}}_{-}
+\frac{1}{2} \bar{G}_{\pi}^{(1)}\,
( \tau_{+}^{z}\, {\vc{\sigma}}_{-} +\tau_{-}^{z}\,{\vc{\sigma}}_{+} )
\nonumber\\
&&\hspace{8em}
+\bar{G}_{\pi}^{(2)}\, (3\tau_{1}^{z}\tau_{2}^{z}- {\vc{\tau}}_{1}\cdot {\vc{\tau}}_{2})\,{\vc{\sigma}}_{-} 
\bigg\}
\cdot
\frac{ \vc{r}}{r} \,
V(r)
,
\label{eq:CPVhamiltonian}
\end{eqnarray}
where $\vc{r} \equiv \vc{r}_1 - \vc{r}_2$, ${\vc{\sigma}}_{\pm} \equiv {\vc{\sigma}}_1 \pm{\vc{\sigma}}_2$, and ${\vc{\tau}}_{\pm} \equiv {\vc{\tau}}_1 \pm{\vc{\tau}}_2$ denote the relative coordinate, spin, and isospin matrices, respectively, of the nucleons 1 and 2. 
The radial function is given by
$V (r) = 
-\frac{m_\pi}{8\pi m_N} \frac{e^{-m_\pi r }}{r} \left( 1+ \frac{1}{m_\pi r} \right)
$.
The CP-odd $\alpha -N $ and $\alpha -^3$H potentials are obtained by folding \cite{yamanakanuclearedmreview} the CP-odd $N-N$ interaction (\ref{eq:CPVhamiltonian}) with the oscillator constant $b = 1.358$ fm ($\alpha - N$) and $b=1.482$ fm ($\alpha -^3$H).
In Fig. \ref{fig:folding} we display the CP-odd potentials.

\section{The nuclear electric dipole moment}
\label{sec:nuclearedm}

The nuclear EDM generated by the CP-odd nuclear force is given by
\begin{eqnarray}
d_{A}^{\rm (pol)} 
&=&
\sum_{i=1}^{A} \frac{e}{2} 
\langle \, \Phi_J (A) \, |\, (1+\tau_i^z ) \, r_{iz} \, | \, \Phi_J (A) \, \rangle
\nonumber\\
&=&
\bar G_\pi^{(0)}
a_\pi^{(0)}
+\bar G_\pi^{(1)}
a_\pi^{(1)}
+\bar G_\pi^{(2)}
a_\pi^{(2)}
.
\label{eq:polarizationedm}
\end{eqnarray}
where $|\, \Phi_{J} (A)\, \rangle$ is the polarized nuclear state.
We show the results of our calculations in Table \ref{tab:nuclearedm}.

\begin{table}
\caption{
The linear coefficients of Eq. (\ref{eq:polarizationedm}) in unit of $e$ fm for several nuclei.
The symbol ``$-$'' means zero within our framework.
}
\label{tab:nuclearedm}
\begin{center}
\begin{tabular}{llll}
\hline\noalign{\smallskip}
 & $a_\pi^{(0)}$ & $a_\pi^{(1)}$ & $a_\pi^{(2)}$ \\ 
\noalign{\smallskip}\hline\noalign{\smallskip}
$^{2}$H \cite{liu,yamanakanuclearedm}  & $-$ & $0.0145 $ & $-$  \\
$^{3}$He \cite{bsaisou,yamanakanuclearedm}& $0.0059$ & 0.0108 & 0.0168  \\
$^{3}$H \cite{bsaisou,yamanakanuclearedm} & $-0.0059$ & 0.0108 & $-0.0170$  \\
$^{6}$Li \cite{yamanakanuclearedm}& $-$ & 0.022 & $-$  \\
$^{7}$Li \cite{Yamanaka:2018dwa} & $-0.006$ & 0.016 & $-0.017$  \\
$^{9}$Be \cite{yamanakanuclearedm}  & $-$ & $0.014$ & $-$  \\
$^{11}$B \cite{Yamanaka:2018dwa} & $-0.004$ & $0.02$ & $-0.01$  \\
$^{13}$C \cite{c13edm} & $-$ & $-0.0020 $ & $-$  \\
$^{129}$Xe \cite{yoshinaga2} & $7 \times 10^{-5}$ & $7 \times 10^{-5} $ & $4 \times 10^{-4}$  \\
\noalign{\smallskip}\hline
\end{tabular}
\end{center}
\end{table}

\begin{figure}
\begin{center}
\includegraphics[width=0.8\textwidth]{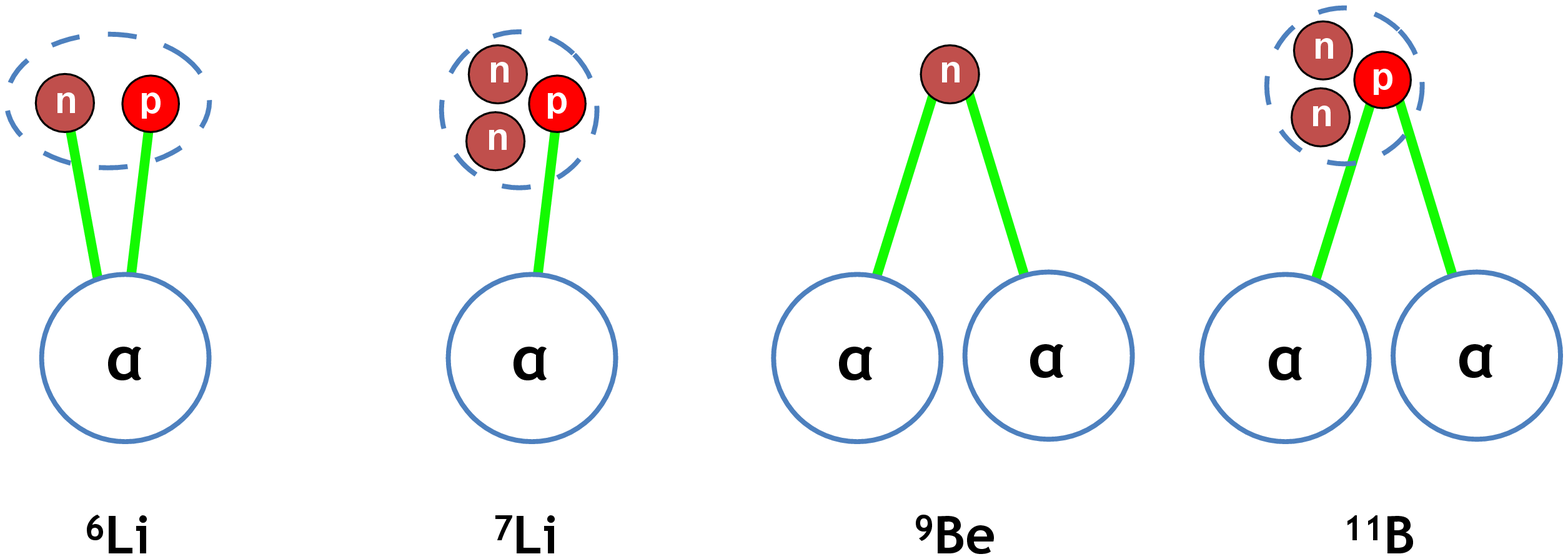}
\caption{
Schematic picture of the counting rule for the EDMs of $^6$Li, $^7$Li, $^9$Li, and $^{11}$B.
}
\label{fig:counting_rule}
\end{center}
\end{figure}

From this result, we can derive an approximate counting rule with the basic components the $^2$H/$^3$H EDM and the CP-odd $\alpha -N \, \mbox{polarization} \sim (0.005 - 0.007)\, \bar G_\pi^{(1)}e$ fm.
We indeed have
\begin{eqnarray}
d_{^{6}{\rm Li}}
& = &
2\times (\alpha -N \, \mbox{polarization})
+ d_{^{2}{\rm H}}
,
\nonumber\\
d_{^{7}{\rm Li}}
& = &
1 \times (\alpha -N \, \mbox{polarization})
+d_{^{3}{\rm H}}
,
\nonumber\\
d_{^{9}{\rm Be}}
& = &
2 \times (\alpha -N \, \mbox{polarization})
,
\nonumber\\
d_{^{11}{\rm B}}
& = &
2 \times (\alpha -N \, \mbox{polarization})
+d_{^{3}{\rm H}}
.
\end{eqnarray}
We display in Fig. \ref{fig:counting_rule} the schematic picture of this rule.
From it, we can predict
\begin{eqnarray}
d_{^{10}{\rm B}}
& \sim &
4\times (\alpha -N \, \mbox{polarization})
+ d_{^{2}{\rm H}}
\sim
0.03 \, \bar G_\pi^{(1)}e \, {\rm fm}
,
\nonumber\\
d_{^{14}{\rm N}}
& \sim &
6 \times (\alpha -N \, \mbox{polarization})
+ d_{^{2}{\rm H}}
\sim
0.04 \, \bar G_\pi^{(1)}e \, {\rm fm}
.
\end{eqnarray}

We note that the EDM of $^{13}$C does not respect the counting rule.
This is due to the bad overlap between the structures of opposite parity states \cite{c13edm}.
This suppression is certainly also relevant for $^{15}$N which has a similar level structure.

Going along with the counting rule, we can naively predict that the EDM will increase if the the nucleon number grows thanks to the $\alpha -N$ polarization.
The nuclear EDM will however be suppressed by the destructive interference due to the configuration mixing for heavy nuclei \cite{yoshinaga2} (see the numerical value of the EDM of $^{129}$Xe in Table \ref{tab:nuclearedm}).

\section{Summary}
\label{sec:summary}

In this proceedings contribution, we presented the results of the calculations of the nuclear EDM.
The EDM of light nuclei seems to obey an approximate counting rule, if the nuclear structures of opposite parity states do not significantly differ.
We could predict that the EDMs of $^{10}$B or $^{14}$N are more sensitive than the known ones.
Increasing the number of nucleons will not give us a sensitive nucleus to the CP violation, since the destructive effect due to the configuration mixing will become important.
Light nuclei seems to be the most suited for the EDM measurement using storage ring experiments.



\begin{thebibliography}{99}
%

\bibitem{engeledmreview}
J. Engel, M. J. Ramsey-Musolf, and U. van Kolck, Prog. Part. Nucl. Phys. {\bf 71}, 21 (2013).

\bibitem{yamanakabook}
N.~Yamanaka,
  ``Analysis of the Electric Dipole Moment in the R-parity Violating Supersymmetric Standard Model,''
Springer, Berlin Germany (2014).
  doi:10.1007/978-4-431-54544-6
  %%CITATION = doi:10.1007/978-4-431-54544-6;%%

\bibitem{atomicedmreview}
N. Yamanaka, B. Sahoo, N. Yoshinaga, T. Sato, K. Asahi, and B. Das, Eur. Phys. J. A {\bf 53}, 54 (2017).

\bibitem{chuppreview}
T. E. Chupp, P. Fierlinger, M. J. Ramsey-Musolf and J. T. Singh, arXiv:1710.02504 [physics.atom-ph].

\bibitem{pospelovckm}
M. Pospelov and A. Ritz, Phys. Rev. D {\bf 89}, 056006 (2014).

\bibitem{seng}
C.-Y. Seng, Phys. Rev. C {\bf 91}, 025502 (2015).

\bibitem{smnuclearedm}
N. Yamanaka and E. Hiyama, JHEP {\bf 02}, 067 (2016).

\bibitem{smdeuteronedm}
N. Yamanaka, Nucl. Phys. A {\bf 963}, 33 (2017).

\bibitem{sm9beedm}
J. Lee, N. Yamanaka, and E. Hiyama, arXiv:1811.00329 [nucl-th].

\bibitem{jedi}
JEDI Collaboration, Phys. Rev. Lett. {\bf 117}, 054801 (2016); Phys. Rev. Lett. {\bf 119}, 014801 (2017).

\bibitem{yamanakanuclearedmreview}
N. Yamanaka, Int. J. Mod. Phys. E {\bf 26}, 1730002 (2017).

\bibitem{devriesedmreview}
J. de Vries and U.-G. Mei{\ss}ner, Int. J. Mod. Phys. E {\bf 25}, 1641008 (2016).

\bibitem{av18}
R. B. Wiringa, V. G. J. Stoks, and R. Schiavilla, Phys. Rev. C {\bf 51}, 38 (1995).

\bibitem{kanada}
H. Kanada, T. Kaneko, S. Nagata, and M. Morikazu, Prog. Theor. Phys. {\bf 61}, 1327 (1979).

\bibitem{hasegawa}
A. Hasegawa and S. Nagata, Prog. Theor. Phys. {\bf 45}, 1786 (1971).

\bibitem{nishioka}
H. Nishioka, S. Saito, and M. Yasuno, Prog. Theor. Phys. {\bf 62}, 424 (1979).

\bibitem{schmid}
E. W. Schmid and K. Wildermuth, Nucl. Phys. {\bf 26}, 463 (1961).

\bibitem{yamada11b}
T. Yamada and Y. Funaki, Phys. Rev. C {\bf 82}, 064315 (2010).

\bibitem{yamada13c}
T. Yamada and Y. Funaki, Phys. Rev. C {\bf 92}, 034326 (2015).

\bibitem{saito}
S. Saito, Prog. Theor. Phys. {\bf 40}, 893 (1968); Prog. Theor. Phys. {\bf 41}, 705 (1969).

\bibitem{pvcpvhamiltonian3}
I. S. Towner and A. C. Hayes, Phys. Rev. C {\bf 49}, 2391 (1994).

\bibitem{liu}
C.-P. Liu and R. G. E. Timmermans, Phys. Rev. C {\bf 70}, 055501 (2004).

\bibitem{yamanakanuclearedm}
N. Yamanaka and E. Hiyama, Phys. Rev. C {\bf 91}, 054005 (2015).

\bibitem{bsaisou}
J. Bsaisou, J. de Vries, C. Hanhart, S. Liebig, U.-G. Mei{\ss}ner, D. Minossi, A. Nogga, and A. Wirzba, JHEP {\bf 03}, 104 (2015) [Erratum ibid. {\bf 05}, 083 (2015)].

\bibitem{Yamanaka:2018dwa}
N. Yamanaka, Hyperfine Interact.\  {\bf 239} (2018) 35 [arXiv:1805.05982 [nucl-th]].

\bibitem{c13edm}
N. Yamanaka, T. Yamada, E. Hiyama, and Y. Funaki, Phys. Rev. C {\bf 95}, 065503 (2017).

\bibitem{yoshinaga2}
N. Yoshinaga, K. Higashiyama, R. Arai, and E. Teruya, Phys. Rev. C {\bf 89}, 045501 (2014).

\end{thebibliography}
\end{document}